\documentclass[sigconf]{acmart}
\AtBeginDocument{%
  }

\setcopyright{acmlicensed}
\copyrightyear{2018}
\acmYear{2018}
\acmDOI{XXXXXXX.XXXXXXX}
\acmConference[Conference acronym 'XX]{Make sure to enter the correct
  conference title from your rights confirmation email}{June 03--05,
  2018}{Woodstock, NY}
\acmISBN{978-1-4503-XXXX-X/2018/06}

\begin{document}

\title{Loop Invariant Generation: A Hybrid Framework of Reasoning optimised LLMs and SMT Solvers}
\author{Varun Bharti}
\authornote{Both authors contributed equally to this research}
\affiliation{%
  \institution{IIIT Delhi}
  \city{Delhi}
  \country{India}}
\email{varun22562@iiitd.ac.in}

\author{Shashwat Jha}
\authornotemark[1]
\affiliation{%
  \institution{IIIT Delhi}
  \city{Delhi}
  \country{India}}
\email{shashwat22472@iiitd.ac.in}
\author{Dhruv Kumar}
\affiliation{%
  \institution{BITS Pilani}
  \city{Pilani}
  \country{India}}
\email{dhruv.kumar@pilani.bits-pilani.ac.in}
\author{Pankaj Jalote}
\affiliation{%
  \institution{IIIT Delhi}
  \city{Delhi}
  \country{India}}
\email{jalote@iiitd.ac.in}
\begin{abstract}
Loop invariants are essential for proving the correctness of programs with loops. Developing loop invariants is challenging, and fully automatic synthesis cannot be guaranteed for arbitrary programs. Some approaches have been proposed to synthesize loop invariants using symbolic techniques and more recently using neural approaches. These approaches are able to correctly synthesize loop invariants only for subsets of standard benchmarks. In this work, we investigate whether modern, reasoning-optimized large language models can do better. We integrate OpenAI’s O1, O1-mini, and O3-mini into a tightly coupled generate-and-check pipeline with the Z3 SMT solver, using solver counterexamples to iteratively guide invariant refinement. We use Code2Inv benchmark, which provides C programs along with their formal preconditions and postconditions. On this benchmark of 133 tasks, our framework achieves 100\% coverage (133/133), outperforming the previous best of 107/133, while requiring only 1–2 model proposals per instance and 14–55 seconds of wall-clock time. These results demonstrate that LLMs possess latent logical reasoning capabilities which can help automate loop invariant synthesis. While our experiments target C-specific programs, this approach should be generalizable to other imperative languages. 

\end{abstract}



\maketitle

\section{Introduction}\label{sec:intro}
Loop invariants are logical assertions that characterize exactly those program states that hold both immediately before and immediately after each iteration of a loop.  Concretely, consider the annotated fragment

\[
  \{P\}\;\mathbf{while}\;(B)\;\{S\};\;\{Q\}
\]

where \(P\) is the precondition, \(B\) the loop guard, \(S\) the loop body, and \(Q\) the postcondition.  An invariant \(I\) must satisfy three finite checks:  
\[
  P \;\implies\; I,
  \quad
  I \land B \;\implies\; I',
  \quad
  I \land \neg B \;\implies\; Q,
\]
where \(I'\) denotes \(I\) interpreted over the state resulting from executing \(S\).  By discharging these implications, deductive verifiers avoid reasoning about an unbounded number of loop iterations, making loop invariants indispensable for proving correctness properties automatically.  

Despite this, generating correct inductive invariants is a major bottleneck. Invariant synthesis is undecidable in general, and manual annotation is tedious and error-prone. While automated techniques exist, key challenges remain. Static methods like abstract interpretation overapproximate reachable states via numeric domains (intervals, octagons, polyhedra)~\cite{cousot1977abstract, blanchet2003static}, but require expert domain choices and struggle with non-linear or modular patterns. Counterexample guided abstraction refinement ( CEGAR )refines abstractions~\cite{clarke2003counterexample}. Template-based solvers assume invariant shapes and solve for parameters~\cite{gulwani2008template}, failing outside predefined templates. Interpolation techniques extract invariants from failed proofs~\cite{mcmillan2003interpolation}, but depend on finding such proofs. Dynamic tools like Daikon mine likely invariants from traces~\cite{ernst2007daikon}, but require formal validation and can miss edge cases.

\begin{figure*}[t]
    \centering
    \includegraphics[width=\textwidth]{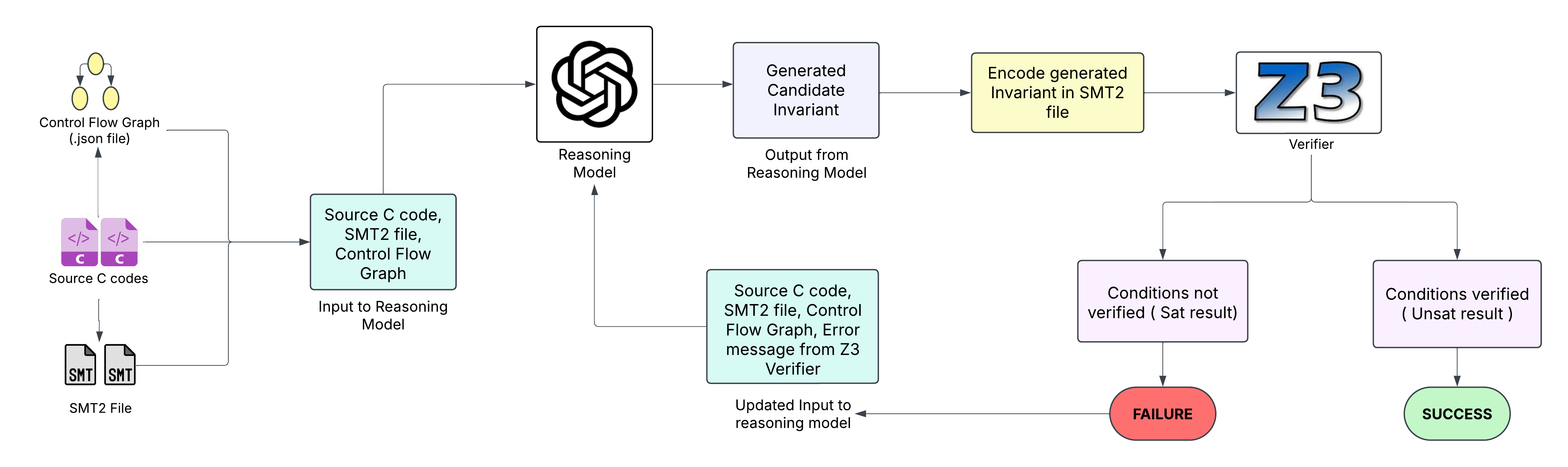}
    \caption{Flow diagram of the generate–and–check loop used for invariant synthesis.}
    \label{fig:invariant-pipeline}
\end{figure*} 

Machine learning approaches aim to learn patterns from data. Code2Inv treats invariant synthesis as a reinforcement learning task, training a policy network to interact with an SMT solver, solving 92/133 benchmarks~\cite{si2018code2inv}. CLN2Inv and its nonlinear variants learn differentiable representations, capturing polynomial and complex relations~\cite{ryan2020cln2inv, yao2020nonlinear}, but are data-dependent and generalize poorly to unseen structures.

Recent work has explored using large language models for loop invariant generation \cite{pei2023can}, \cite{wu2024lemur}, with LEMUR~\cite{wu2024lemur} emerging as a particularly influential framework. LEMUR prompts GPT-3.5\cite{gpt3.5} and GPT-4\cite{gpt4} with loop code and associated verification conditions to synthesize candidate invariants. A key innovation is its feedback mechanism: when a generated invariant fails verification, counterexamples are extracted and incorporated into the next prompt to guide refinement. This generate–check–repair loop enables LEMUR to achieve strong performance, solving 107 out of 133 benchmarks in the Code2Inv dataset~\cite{wu2024lemur}.

Our framework adopts the principle of prompt refinement guided by counterexamples, but extends it beyond LEMUR’s C-specific toolchain. While LEMUR integrates with C‑specific SMT‑based model checkers such as CBMC, ESBMC, and UAutomizer, our approach leverages Z3, an SMT-LIB–compliant solver agnostic to the source language. This design enables broader applicability: any imperative language can be supported by emitting suitable SMT encodings.

To motivate the generate–and–check framework, we first show how any candidate invariant \(I\) can be validated by reducing it to a sequence of SMT-LIB satisfiability queries. Given precondition $P$, loop guard $B$, and postcondition $Q$, we assert the negation of each verification condition in SMT-LIB.  Concretely, we generate three assertions:
\begin{align*}
  &\texttt{(assert (not (=> P I)))}\\
  &\texttt{(assert (not (=> (and I B) I')))}\\
  &\texttt{(assert (not (=> (and I (not B)) Q)))}\\
  &\texttt{followed by \texttt{(check-sat)}}.
\end{align*}

This reduction to a finite sequence of SMT queries naturally enables the integration of LLM-generated invariant candidates with automatic, solver-driven validation and iterative refinement.

In this work, we present our approach to loop invariant generation that tightly integrates powerful reasoning based language models with language neutral automated verifiers. Our framework iteratively alternates between two core phases: (1) An inference phase, where the model synthesizes a candidate invariant from the program text and specification, and (2) a formal verification phase, where an SMT solver checks the candidate against the initialization, inductive, and postcondition obligations.  Whenever the solver detects a failure, it produces a concrete counterexample that is fed back into the next inference prompt, guiding the model to refine or strengthen its proposal.  This generate-and-check paradigm ensures that every accepted invariant is formally proven correct, while exploiting the model’s capacity for semantic insight and pattern recognition. 
 We evaluate the framework on Code2Inv benchmark~\cite{si2018code2inv},achieving 100\% coverage on all cases. The performance gains arise primarily from selecting reasoning optimized LMs rather than from the verifier itself. The models utilized are :
\begin{itemize}
  \item \textbf{O1-mini (2024)}\cite{gpto1mini}: a compact, fast model optimized for multi-step deduction under tight latency constraints.
  \item \textbf{O3-mini (2025)}\cite{gpto3mini}: a next generation variant with expanded context and enhanced logical consistency.
  \item \textbf{O1 (2024)}\cite{gpto1}: the flagship reasoning model, offering the highest single-shot accuracy at the cost of larger per-query latency.
\end{itemize}

\section{Proposed Methodology}

Our approach is based on a tightly coupled \emph{generate–and–check} loop between an LLM and an SMT solver.  Beginning with a C program annotated by precondition \(P\), loop guard \(B\), and postcondition \(Q\), we first query the LLM to propose a candidate invariant \(I\) in SMT‑LIB syntax.  We then splice \(I\) into an SMT2 template asserting the negations of the three verification conditions introduced in Section~\ref{sec:intro} and invoke Z3. If Z3 returns \texttt{unsat}, no counterexample exists and \(I\) is accepted as a correct inductive invariant. Otherwise, Z3 produces either a concrete counterexample (\texttt{sat}) or a parse error, which we capture and feed back into the LLM as a repair hint.  This process repeats until the invariant verifies or we reach our iteration cap.  The full pipeline is illustrated in Figure~\ref{fig:invariant-pipeline}.  

\subsection{Data Preprocessing}
We begin by parsing each C program using the Code2Inv front‐end \cite{si2018code2inv} to extract:
\begin{itemize}\itemsep0pt
  \item The loop guard \(B\) and control‐flow graph (as JSON).
  \item An SMT2 template with placeholders for the invariant \(I\).
\end{itemize}
This stage is purely syntactic and yields the inputs required by both the LLM and the verifier.

\noindent Although our prototype uses a C‐specific front‐end to generate the CFG and SMT2 template, the overall pipeline is language‐agnostic: any language whose code can be parsed into a control‐flow graph and a corresponding SMT‑LIB formula can plug in directly.  For instance, Python scripts can be translated to CFGs via tools like PyCFG~\cite{pycfg} or the PythonTA `[cfg]` extension~\cite{pythonta_cfg}, Java bytecode CFGs can be extracted with Soot~\cite{soot}, and SMT2 templates can be emitted by OpenJML~\cite{openjml}.  More generally, multi‐language frameworks such as SMACK ingest LLVM bitcode (from C, C++, Rust, even Python via llvmlite) and produce Boogie or SMT2 directly~\cite{smack}.  Adopting these common intermediate representations is substantially easier than building a bespoke model checker or verifier for each new programming language.  

\subsection{Reasoning Model}
Our LLM component is invoked twice per iteration: once to propose an initial invariant and again to repair it when needed.  In both calls, we present:
\begin{itemize}\itemsep0pt
  \item The source C code, its control‐flow graph, and the SMT2 template.
  \item For the repair call, we additionally include a concrete counterexample given by Z3
  \item An instruction
\end{itemize}

\paragraph{Initialization Call}
The first LLM query asks for a fresh candidate invariant \(I\).  Upon receipt, we splice \(I\) into the SMT2 template to form a concrete verification problem. This SMT2 file is sent to Z3 verifier for verification.

\paragraph{Repair Call}
If any of the three SMT checks fails (\texttt{sat} or parse error), we construct a second prompt that includes:
\begin{itemize}\itemsep0pt
  \item The original SMT2 template and all previously proposed invariants.
  \item Either the Z3 counterexample model or the SMT parse error message.
  \item An instruction to “Refine the invariant to rule out this counterexample/error.”
\end{itemize}
The LLM’s output replaces the old invariant, and we loop back to verification.

\subsection{Verifier: SMT-based Validation with Z3}

We use Z3~\cite{demoura2008z3} to discharge the three invariant proof obligations automatically. Z3 ingests formulas in SMT‑LIB format under the theory of linear integer arithmetic and returns either:
\begin{itemize}\itemsep0pt
  \item \texttt{unsat}, indicating no counterexample exists (the invariant holds), or
  \item \texttt{sat}, providing a concrete model that violates one of the checks, or
\end{itemize}

To validate a candidate invariant \(I\) against precondition \(P\), guard \(B\), and postcondition \(Q\), we emit three negated implication assertions. The Z3 verifier then:
\begin{enumerate}\itemsep0pt
  \item Parses and checks each of the three assertions.
  \item If all return \texttt{unsat}, the candidate \(I\) is confirmed as a valid inductive invariant.
  \item Otherwise, a \texttt{sat} result yields a counterexample model, and a parse error yields an SMT‑LIB syntax message, both of which are forwarded to the repair prompt.
\end{enumerate}

\textbf{Note on SMT Errors:} If Z3 does not parse the generated invariant due to invalid SMT-LIB syntax, we capture the error message string and treat it as a special 'counterexample' in the next stage.

\medskip
\noindent\textbf{Example 1 (SAT Repair):}  
Consider the \texttt{122.c} benchmark, where the LLM initially proposes the following invariant:
\[
  \mathtt{(=\,sn\,(-\,i\,1))}.
\]
When we plug this into our SMT template and invoke Z3, the solver reports \texttt{sat} and returns the following counterexample:
\[
i = 0,\quad \text{size} = -2,\quad sn = -1.
\]
Since \(\neg B\) holds at exit (\(0 > -2\)), and the candidate invariant \(sn = i - 1\) is also satisfied \((-1 = 0 - 1)\), both the initialization and inductiveness checks pass. However, the postcondition requires \(\mathit{sn} = 0\) upon loop exit, which is violated here (\(-1 \neq 0\)). Z3 therefore reports “sat” and returns this concrete model as a counterexample.

This counterexample is fed back to the LLM in the repair prompt.  In response, the model synthesizes a strengthened invariant, for example  
\[
  \mathtt{(and\,(>=\,i\,1)\,(=\,sn\,(-\,i\,1))\,(<=\,i\,(+\;size\;1)))},
\]
which Z3 subsequently verifies as \texttt{unsat} on all three checks, confirming it is indeed a correct inductive invariant.

\medskip
\noindent\textbf{Example 2 (SMT Error Repair):}  
The LLM may also generate invariants with incorrect SMT-LIB syntax, in which case Z3 returns a parsing error.
\medskip

This tightly‐coupled pipeline of generation, formal checking, and counterexample driven repair converges in just 1-5 iterations across 133 benchmarks, as shown in the next section

\section{Experimental Design and Results}

\subsection{Experimental Design}
To validate our approach, we ran experiments on the Code2Inv benchmark suite which comprises 133 C programs each containing a loop.~\cite{si2018code2inv}.  We evaluated three configurations of reasoning-optimized LLMs: O1-mini, O3-mini, and O1 via OpenAI’s API at temperature 0 to ensure deterministic outputs and leveraging their large context windows for precise code reasoning.  

Based on observations from prior work that LLM based loops can stagnate after many repair iterations~\cite{wu2024lemur, kamath2023finding}, we experimented with various iteration limits and chose \(N=5\) as a balance between refinement capacity and avoiding redundant proposals.  In every run, we used Z3 v4.8.17 to discharge the initialization, inductiveness, and postcondition checks, with each query subject to a 5 s timeout~\cite{demoura2008z3}.

We recorded three key metrics per program: (1) Mean wall-clock time (LLM latency + SMT solving), (2) number of LLM proposals (iterations) until success, and (3) Mean memory footprint of the verification engine. To contextualize our results, we compared against four established baselines:
\begin{itemize}
  \item ESBMC (k-induction model checker): solved 68 of 133 tasks~\cite{gadelha2018esbmc}.
  \item Code2Inv (deep RL invariant synthesis): solved 92 of 133 tasks~\cite{si2018code2inv}.
  \item LEMUR–GPT-3.5: integrated GPT-3.5 with C -specific SMT- based model checker to solve 103 of 133~\cite{wu2024lemur}.
  \item LEMUR–GPT-4: the same framework with GPT-4, solving 107 of 133~\cite{wu2024lemur}.
\end{itemize}

\subsection{Results and Analysis}
Table~\ref{tab:perf} summarizes overall performance: all three LLM configurations achieved perfect coverage, automatically synthesizing a correct inductive invariant for every benchmark.  The table also reports average wall-clock time and the mean number of iterations (LLM proposals) per program.

\begin{table}[ht]
  \caption{Performance on Code2Inv}\label{tab:perf}
  \centering
  \begin{tabular}{l c c c}
    \hline
    \textbf{Method} & \textbf{Solved} & \textbf{Time (s)} & \textbf{Iters}\\
    \hline
    ESBMC            & 68   & 0.34   & –   \\
    Code2Inv         & 92   & –      & –   \\
    LEMUR-GPT3.5     & 103  & 35.6   & 8.6 \\
    LEMUR-GPT4       & 107  & 32.9   & 4.7 \\
    \hline
    O1-mini + Z3 (ours)   & \textbf{133} & 14.5 & 1.04 \\
    O3-mini + Z3 (ours)   & \textbf{133} & 25.9 & 1.37 \\
    O1 + Z3  (ours)     & \textbf{133} & 55.5 & 1.00 \\
    \hline
  \end{tabular}
\end{table}

\begin{table}[ht]
  \caption{Average Time and Memory Usage per Model}\label{tab:resources}
  \centering
  \begin{tabular}{l c c}
    \hline
    \textbf{Model}      & \textbf{Avg Time (s)} & \textbf{Avg Memory (MB)}\\
    \hline
    O1-mini + Z3 (ours)      & 14.52                 & 0.150                    \\
    O3-mini + Z3 (ours)       & 25.89                 & 0.345                    \\
    O1 + Z3  (ours)           & 55.49                 & 0.158                    \\
    \hline
  \end{tabular}
\end{table}

\subsubsection{Quantitative Analysis :}
Our reasoning optimized models not only achieve 100\% coverage, but do so with markedly lower iteration counts and resource demands than prior LLM-based frameworks.  As Table~\ref{tab:perf} shows, LEMUR-GPT4 averaged 4.7 proposals per instance, whereas O1 and O1-mini succeeded in a single shot  (mean \(\approx\!1.0\) proposals) and O3-mini required only 1.37 proposals on average.  This translates directly into reduced wall-clock time: O1-mini completes proofs in about 14.5 s on average, roughly half the 32.9 s reported for GPT-4, and O3-mini finishes in under 26 s.  Table~\ref{tab:resources} further breaks down memory usage: despite solving all tasks in one iteration, O1’s larger model size yields a slightly higher memory footprint than O1-mini, while O3-mini’s iterative refinements inflate its prompt and solver state to about 0.34 MB per instance.

The per‐iteration performance of each reasoning model is summarized in Table~\ref{tab:iterations}.  O1 converges immediately, solving all 133 benchmarks on its first proposal.  O1‑mini succeeds on 128 problems in one shot and completes the remaining 5 on its second attempt.  O3‑mini exhibits a slightly longer tail: it solves 95 instances in the first round, 28 in the second, 8 in the third, and 2 in the fourth, reaching full coverage by iteration four.

\begin{table}[ht]
  \caption{Benchmarks Solved vs.\ Iteration Count}\label{tab:iterations}
  \centering
  \begin{tabular}{l c c c c c}
    \hline
    \textbf{Model} & \textbf{Iter 1} & \textbf{Iter 2} & \textbf{Iter 3} & \textbf{Iter 4} & \textbf{Iter 5} \\
    \hline
    O1                 & 133            & 0              & 0              & 0              & 0              \\
    O1‑mini            & 128            & 5              & 0              & 0              & 0              \\
    O3‑mini            & 95             & 28             & 8              & 2              & 0              \\
    \hline
  \end{tabular}
\end{table}
This consistent behavior across models suggests that, once guided by Z3 counterexamples, even compact LLMs can rapidly converge to correct invariants with minimal iteration overhead.

\subsubsection{Qualitative Analysis :}
We also noted that across a representative subset of benchmarks, our models consistently generated concise, human readable invariants.  In simple loops with linear updates (e.g., incrementing \(x\) until \(n\)), the LLMs produced the canonical assertion 
\[
  0 \le x \le n,
\]
and in some cases even inferred equivalent off‐by‐one variants (e.g., \(0 \le x < n+1\)).  For more intricate loops, those combining conditional updates or modular arithmetic, the models output accurate conjunctive and modulus based invariants (e.g., \(x \bmod k = r\) and \(0 \le y \le m \land x = 2y\)), patterns that typically require specialized abstract domains or manual insight.

The feedback driven repair loop proved pivotal for corner cases.  When a first proposal omitted a necessary constraint (such as a lower bound), Z3 returned a concrete counterexample, which the LLM used to refine its invariant in the subsequent prompt.  This targeted correction often succeeded in one additional iteration.  All generated invariants were formally verified by Z3 and remained easily interpretable, indicating that our LLM–SMT pipeline can produce both correct and transparent proofs suitable for integration into development workflows.  We note that ranking and selecting among multiple invariants is an orthogonal concern, addressed by recent work on ranking LLM candidates to improve selection quality~\cite{chakraborty2023ranking}.

\section{Conclusion and Future Work}

Loop invariants lie at the heart of deductive program verification, yet their automatic generation remains a longstanding challenge due to undecidability and the complexity of real world loops. Early symbolic methods and dynamic mining techniques provided partial solutions, and recent work has shown that general purpose LLMs can propose invariants with some success.

In this work, we evaluated the capabilities of state-of-the-art reasoning optimized LLMs (O1, O1-mini, O3-mini) for generating loop invariants, when integrated with Z3 based validation and counterexample generation. We demonstrated that these models can synthesize inductive invariants effectively. Our framework was evaluated on the standard Code2Inv suite and achieved 100 \% coverage requiring an average of 1-2 proposals and under a minute of wall-clock time per instance. By combining reasoning optimized LLMs with rigorous solver feedback, we demonstrate that these models can perform logically sound inference in formal verification tasks, a capability that challenges the status quo and opens new directions for AI-driven software engineering.

We plan to extend this methodology to richer program constructs and other imperative language programs, including nested loops, pointer manipulations, and heap-allocated data structures, to explore the boundaries of LLM-based reasoning. We aim to investigate ensemble strategies, leveraging multiple models and prompt variations to further reduce iteration counts and enhance robustness. Our long term goal is to move beyond loop invariants toward end-to-end proof synthesis, where LLMs generate not only invariants but also intermediate assertions and full proof scripts, bringing us closer to fully automated, AI-driven deductive verification for a wide range of programming languages.

\bibliographystyle{ACM-Reference-Format}
\bibliography{base}

\appendix

\end{document}